\documentclass[letterpaper, 10pt, twocolumn]{article}
\usepackage[hmargin=1.9cm,vmargin=2.5cm]{geometry}
 
\usepackage[utf8]{inputenc}
\usepackage[T1]{fontenc}
\usepackage[english]{babel}
\usepackage{microtype}
\usepackage{hyperref}
\usepackage{url}

\usepackage{changes}
\usepackage{lipsum}
\setcommentmarkup{\footnotesize\itshape\color{red}-- #1 --}

\usepackage{amsmath}
\usepackage{amsfonts}
\usepackage{amssymb}
\usepackage{mathtools}
\usepackage{lmodern}
\usepackage{upgreek}
\usepackage{ragged2e}
\usepackage{cuted}
\usepackage[version=3]{mhchem}
\usepackage[retain-explicit-plus, detect-all]{siunitx}
\usepackage{circledsteps}

\usepackage{titling}
\usepackage{titlesec}
\usepackage[affil-it]{authblk}

\pretitle{\begin{center}\large\bfseries}
\posttitle{\par\end{center}}
\preauthor{\begin{center}
\lineskip 0.25em%
\begin{tabular}[t]{c}}
\postauthor{\end{tabular}\par\end{center}}
\predate{\begin{center}\small}
\postdate{\par\end{center}\vskip-1cm}
\date{}

\titleformat{\section}{\normalfont\large\bfseries}{}{0pt}{}
\titlespacing{\section}{0pt}{5pt plus 5pt minus 3pt}{2pt}

\titleformat{\subsection}[runin]{\normalfont\bfseries}{}{0pt}{}[\mbox{ -- }]
\titlespacing{\subsection}{0pt}{4pt}{4pt plus 2pt minus 1pt}

\titleformat{\subsubsection}[runin]{\normalfont\itshape}{}{0pt}{}[: ]
\titlespacing{\subsubsection}{0pt}{3pt minus 3pt}{4pt plus 2pt minus 1pt}

\makeatletter
\renewcommand{\fnum@figure}{\textbf{Figure~\thefigure}}
\makeatother

\usepackage[%
    backend=biber,
    style=nature,
    alldates=year,
    doi=false,
    isbn=false,
    url=false,
    eprint=true
]{biblatex}

\DeclareSourcemap{
  \maps{
    \map{
      \step[fieldset=note, null]
      \step[fieldset=address, null]
      \step[fieldset=location, null]
      \step[fieldset=language, null]
    }
  }
}

\newbibmacro{string+doiurl}[1]{%
	\iffieldundef{doi}
	{\iffieldundef{url}
		{#1}
		{\href{\thefield{url}}{#1}}}
	{\href{https://doi.org/\thefield{doi}}{#1}}}
 
\DeclareFieldFormat{title}{\usebibmacro{string+doiurl}{\mkbibemph{#1}}}

\usepackage{enumitem}
\usepackage{placeins}
\usepackage{subcaption}
\DeclareCaptionLabelFormat{paren}{(#2)}
\DeclareCaptionJustification{justified}{\justifying}
\usepackage{epstopdf}
\usepackage{graphicx}
\graphicspath{{./Figs/}}

\usepackage{varwidth}
\usepackage{xcolor}
\definecolor{mossgreen}{rgb}{0.68, 0.87, 0.68}
\definecolor{mustard}{rgb}{1.0, 0.86, 0.35}
\definecolor{frenchblue}{rgb}{0.0, 0.45, 0.73}
\definecolor{lightskyblue}{rgb}{0.53, 0.81, 0.98}
\definecolor{purpleheart}{rgb}{0.41, 0.21, 0.61}
\definecolor{jasper}{rgb}{0.84, 0.23, 0.24}

\usepackage{tikz}
\usetikzlibrary{calc,fit,shapes,arrows,positioning,shadows,shapes.symbols}

\tikzset{block/.style = {
		rectangle, draw, fill=frenchblue!25, rounded corners, minimum height=3em,inner sep=5pt,%
		drop shadow,
		execute at begin node={\begin{varwidth}{#1}\centering},%
			execute at end node={\end{varwidth}}
	},
	block/.default={9em}
}
\tikzstyle{decision} = [diamond, aspect=1.5, draw, fill=jasper!30, 
drop shadow,
text width=4em, text badly centered, inner sep=0pt,outer sep = -.5pt]
\tikzstyle{cloud} = [draw, ellipse,fill=mustard,
drop shadow,
minimum height=2em]
\tikzstyle{line} = [draw, -stealth, line width=1pt]

\usepackage{cleveref}

\newcommand{\figsublabel}[1]{\protect\phantomsubcaption\label{#1}\protect\bfseries\textsf{\subref*{#1}}}%

\usepackage{xparse} %
\usepackage{xspace}

\ExplSyntaxOn
\prop_new:N \l_pollard_a_prop
\cs_new:Npn \pollard_wrappera:nn #1#2
{%
	\textbf{\subref*{#1}}\nobreakspace\ignorespaces#2\unskip\ignorespacesafterend\hspace{4pt plus 2pt minus 2pt}%
}

\NewDocumentCommand { \addsubcapLbl } {sO{black}m m m m }
{%
	\IfBooleanTF#1
	{\def \fillingArg{}}
	{\def \fillingArg{fill=white}}
	\protect\node[rectangle, text=#2, anchor=north~west,style/.expanded=\fillingArg] at (#4,#5) {\figsublabel{#3}};
	\prop_gput:Nnn \l_pollard_a_prop {#3} {#6}
}
\NewDocumentCommand { \addsubcap } { m m }
{%
	{\phantomsubcaption\label{#1}}%
	\prop_put:Nnn \l_pollard_a_prop { #1 } { #2 }%
}

\NewDocumentCommand { \processCaptions } { }
{
	\prop_map_function:NN \l_pollard_a_prop \pollard_wrappera:nn
	\prop_gclear_new:N  \l_pollard_a_prop
}
\ExplSyntaxOff

\begin{document}
\begin{refsection}

\author[1]{Van Doan Le}

\author[1]{Julien Fatome}

\author[1]{Erwan Lucas\thanks{erwan.lucas@ube.fr}}

\affil[1]{Laboratoire Interdisciplinaire Carnot de Bourgogne ICB UMR 6303, Université Bourgogne Europe, CNRS, 21000 Dijon, France}

\title{Direct Observation of Switching-Wave Dynamics\\in 20 GHz Normal Dispersion Spiral Microresonators}

\maketitle
\thispagestyle{empty}
\noindent\textbf{\boldmath
	Integrated Kerr frequency combs are powerful tools for microwave photonics, spectroscopy, and optical communications. While traditional architectures rely on the anomalous-dispersion regime, this typically requires thick, highly strained silicon nitride layers that complicate standard CMOS-foundry fabrication.
	Thinner layers circumvent these fabrication constraints but yield normal-dispersion microcombs that generally require specific trigger mechanisms, such as deterministic seeding, due to the absence of spontaneous modulational instability.
	Here, we demonstrate the generation of a \qty{20}{\GHz} microcomb in the normal-dispersion regime, driven by a synchronized dual-pump scheme via electro-optic sidebands modulation.
	The platform leverages an Archimedean spiral geometry on a foundry-compatible, 350 nm-thin SiN platform.
	By employing adiabatic curvature engineering, the resonator balances a compact footprint with an intrinsic quality factor exceeding $7 \times 10^6$, while effectively eliminating avoided mode crossings with higher order modes.
	This strong-dispersion architecture yields a low-repetition-rate microcomb featuring high power-per-line and picosecond-scale temporal profile, which enables the direct optical sampling and characterization of the out-coupled switching wave waveforms.
	 The measured dynamics across a range of pump desynchronizations demonstrate excellent agreement with simulations. Our work establishes a scalable, strategy for footprint-efficient normal-dispersion microcomb generation at microwave frequencies, with potential for scaling to other wavelength ranges.
}

\section*{Introduction}

Optical Kerr frequency combs (microcombs) are broadband sources consisting of evenly spaced spectral lines, generated through the interplay of Kerr nonlinearity, dispersion, and dissipation in coherently driven microresonators~\cite{Pasquazi2018MicrocombsNovel,Kippenberg2018}.  

Today, most demonstrations rely on anomalous group-velocity dispersion (GVD), which enables the formation of dissipative Kerr solitons (DKS) pulses~\cite{Leo2010, Herr2013} that support robust and spontaneous coherent comb formation. However, achieving anomalous dispersion typically requires thick (\qty{>700}{\nm}) silicon nitride (\ce{Si3N4}) waveguides, which are more complex to manufacture~\cite{Pfeiffer2016} and are not widely accessible in standard silicon photonics processes.

A second challenge is scaling microresonators to lower free spectral range (FSR) to reach microwave repetition rates. Increasing cavity length reduces the intra-cavity intensity, requiring higher quality factors to sustain nonlinear dynamics. Conventional approaches rely on large-radius ring resonators, whose footprint scales approximately with $1/\mathrm{FSR}^2$~\cite{Liu2020PhotonicMicrowave}. Spiral resonators offer a compact alternative, but curvature discontinuities may induce coupling to higher-order modes (HOMs), resulting in avoided mode crossings that disrupt the comb formation.

We address these challenges by demonstrating normal-dispersion microcomb generation using Archimedean spiral resonators with a \SI{\sim20}{GHz} FSR on a CMOS-compatible \SI{350}{nm}-thick SiN platform.
This geometry suppresses higher-order mode excitation through smooth curvature engineering~\cite{Chen2012GeneralDesign}, and has been previously demonstrated for soliton combs in anomalous GVD~\cite{Ye2022}.

In the normal dispersion regime, spontaneous comb formation emerges on an unstable branch of the CW-driven nonlinear system, which is typically prohibited~\cite{Godey2014}.
Additional ingredients are thus needed to trigger the pattern formation that sustains the comb. 
One option is to work on the modal structure of the resonator, though e.g. avoided mode crossing~\cite{Xue2015}, or photonic bandgap~\cite{Yu2021} to restore a phase matching pathway.
Another approach relies on self-injection locking, where a coherent feedback into the driving laser stabilizes the patterns~\cite{Wang2022c, Lihachev2022}.

A third option is to drive the system with more than a single optical tone. The four wave mixing process is thereby directly seeded and pattern formation becomes thresholdless.
A number of studies, in particular in fiber-based cavities have used a synchronous pulse pattern to drive the cavity~\cite{Macnaughtan2023TemporalCharacteristics,Bunel2024BroadbandKerr}, while others used a pump modulation to achieve this goal~\cite{Liu2022StimulatedGeneration}.
The pattern that arises under these condition rely on the formation of two switching waves (SW)~\cite{Coen1999}, a type of dispersive front that emerge through wave breaking~\cite{Malaguti2014DispersiveWavebreakinga}.
Such combs typically exhibit higher conversion efficiencies compared to bright solitons in the anomalous regime~\cite{Xue2017, Yang2024EfficientMicroresonator}.

Here, we use a dual-pump configuration to seed a modulation pattern in our cavity and perform an accurate study on the spectral and temporal properties of the SW patterns. Temporal and spectral profiles are characterized directly at the bus-waveguide through port without requiring post-amplification nor complex line by line retrieval~\cite{Xue2015}.

\section*{Results and Discussion}
\begin{figure}[t]
	\centering
	\includegraphics[width=\linewidth]{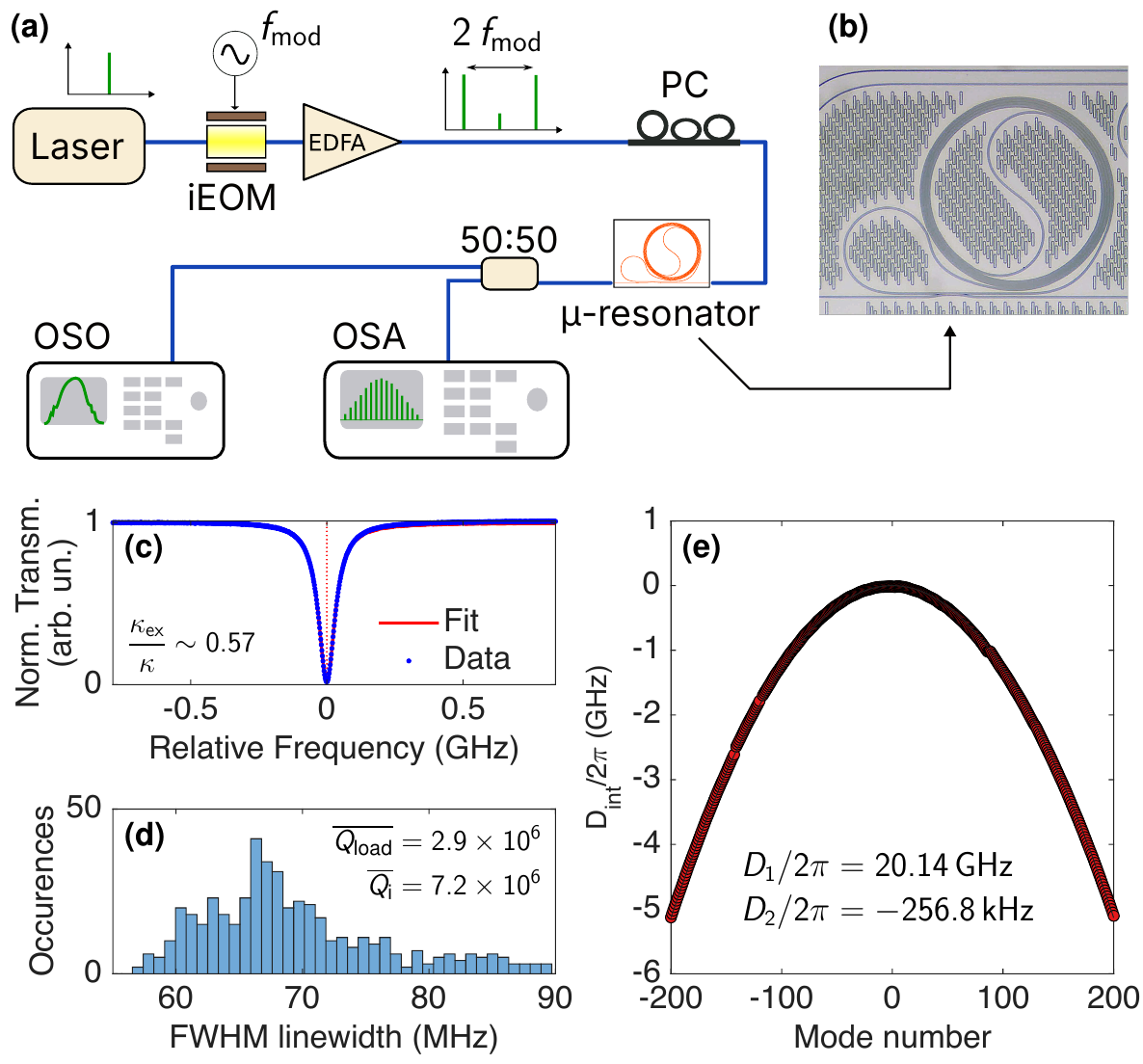}
	\caption{
		(a) Experimental setup for the dual pump comb generation.
		(b) Microscope image of the resonator.
		(c) Measured transmission of one of the pumped resonances.
		(d) Histogram of the recorded linewidth of the resonator.
		(e) Measured integrated dispersion of the spiral resonator respectively.
	}
	\label{fig:1}
\end{figure}

To achieve clean, mode crossing-free normal-dispersion microcombs at low repetition rates, we employ a resonator composed of nested Archimedean spirals connected with smooth curvature bends (see \cref{fig:1}b).
The switching-wave dynamics we demonstrate are not unique to this geometry and can also be realized in standard microring ~\cite{Wang2022c} and racetrack~\cite{Anderson2020a} resonators, as well as other platforms~\cite{Ye2022, Liu2022StimulatedGeneration}.
The resonator consists of a \qty{7.57}{\mm} long waveguide folded into two interleaved spirals with a minimum bend radius of \qty{80}{\micro\meter}. This geometry reduces the footprint 18-fold compared to an equivalent circular ring, from \qty{\sim4.5}{\square\mm} to \qty{0.25}{\square\mm}.
The spiral waveguide features a cross-section of \qtyproduct{1900 x 350}{\nm}. A straight bus--resonator coupler is designed to achieve critical coupling, yielding a coupling ratio of \qty{-23}{\decibel} (\cref{fig:1}c). The facet coupling loss from the lensed fiber to the microresonator is measured to be 3.2 dB/facet.
A calibrated wavelength scan spectroscopy~\cite{DelHaye2009}, indicate a good cavity loading and a mean linewidth value of \qty{68}{\MHz}, corresponding to an estimated intrinsic quality factor of \num{7.2} millions (\cref{fig:1}c,d).
The measurement indicates a FSR of $ D_1/2\pi = \qty{20.14}{\GHz}$ and a strong normal GVD, which induces the quadratic walk-off of the resonances from a uniform FSR. The resonance walk-off is described by the integrated dispersion:
\begin{equation}
	D_{\rm int}(\mu) = \omega_\mu - (\omega_0 + \mu \, D_1) =  \frac{D_2}{2}\mu^2 + \frac{D_3}{6}\mu^3 + \dots ,
\end{equation}
with the relative mode number $\mu$, relative to a reference resonance $\omega_0$.
The measurement yield a factor $D_2/2\pi \sim \qty{-257}{\kHz}$ (\cref{fig:1}e) . The integrated dispersion profile shown in \cref{fig:1}e is smooth and free of observable mode crossings, thanks to the smooth clothoid transitions used at each points of curvature variation. The absence of mode perturbations provide a very clean system that is closest to the idealized models.

\begin{figure*}[ht]
	\centering%
	\includegraphics[width=\linewidth]{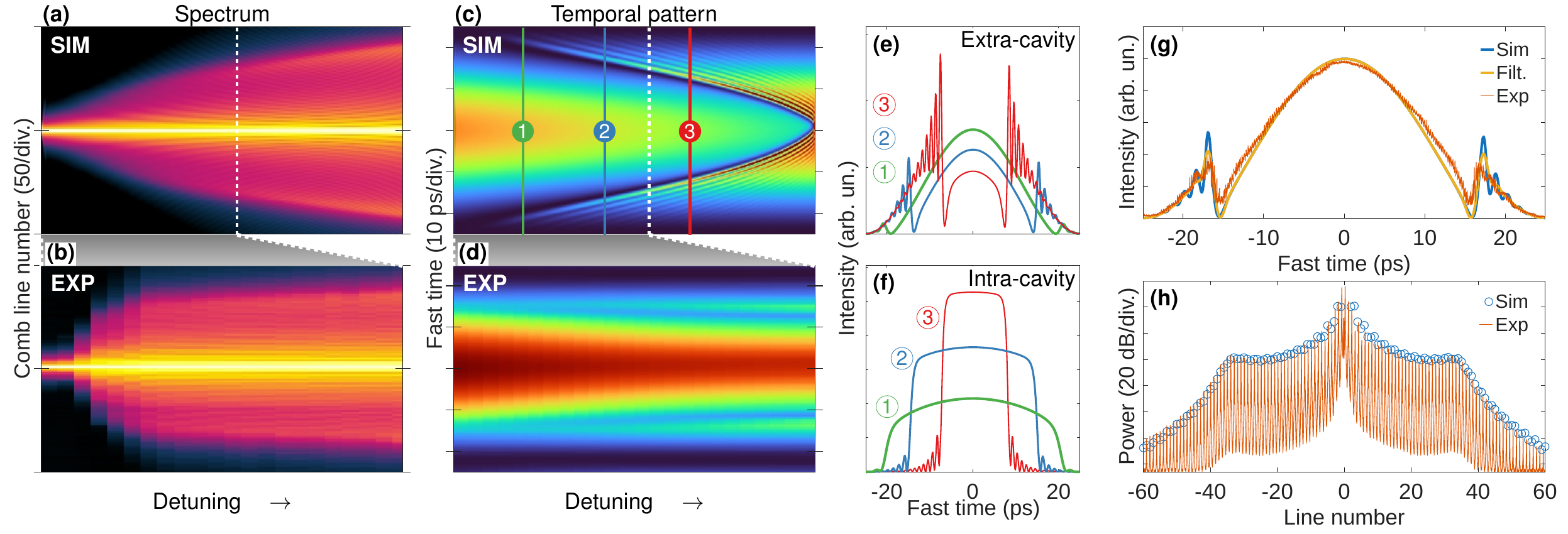}%
	\caption{
		Temporal and spectral evolution with laser detuning for $\Delta\omega_{\rm sync} \sim 0$.
		(a,b,c,d) Simulated and measured spectral envelope (a,b) and temporal intensity (c,d) map as a function of time, while the detuning is gradually increased.The white dashed line in the simulations indicates where the experiment scan stops.
		(e,f) Extra-cavity (e) and intra-cavity (f) temporal intensity patterns at the three different detuning values corresponding to vertical lines in 1,2,3 in (c).
		(g,h) Comparison between the simulated and measured extra-cavity temporal profile (g) and corresponding microcomb (h) corresponding to the white dashed lines in (a, c).
		The filtered curve in (g) represents the simulated results after lowpass filtering.
	}
	\label{fig:2}
\end{figure*}

To generate a frequency comb, we set up an experiment as illustrated in \cref{fig:1}a.
A single frequency continuous-wave laser is converted into dual frequencies by an electro-optic intensity modulator (iEOM) biased at the null point, to suppress the carrier. The modulation frequency is set to $f_{\rm mod} = \text{FSR}/2 =  \qty{10.0436}{\GHz}$, resulting in two sidebands, whose spacing precisely matches the resonator's FSR.%
The sidebands are amplified to an average power of \qty{297}{\mW} at \qty{1550}{\nm}, with an erbium-doped-fiber amplifier (EDFA). This power level was needed to fully excite the formation of the dispersive wave features on the switching waves~\cite{Lottes2021}. The polarization is controlled via a fiber polarization controller (PC) before coupling to the photonic chip via lensed fibers. Comb generation is triggered by gradually scanning the pump laser frequency to simultaneously scan both pump carriers across their respective resonance.
The generated microcomb is characterized at the output of the chip by an optical spectrum analyzer (OSA), while the corresponding  temporal intensity waveform is recorded using an optical sampling oscilloscope (OSO) with a \qty{800}{\GHz} bandwidth.
Both devices are recorded continuously while the laser frequency is tuned to increase the detuning, yielding the maps in \cref{fig:2}b,d.

\begin{figure*}[t]
	\centering%
	\includegraphics[width=\linewidth]{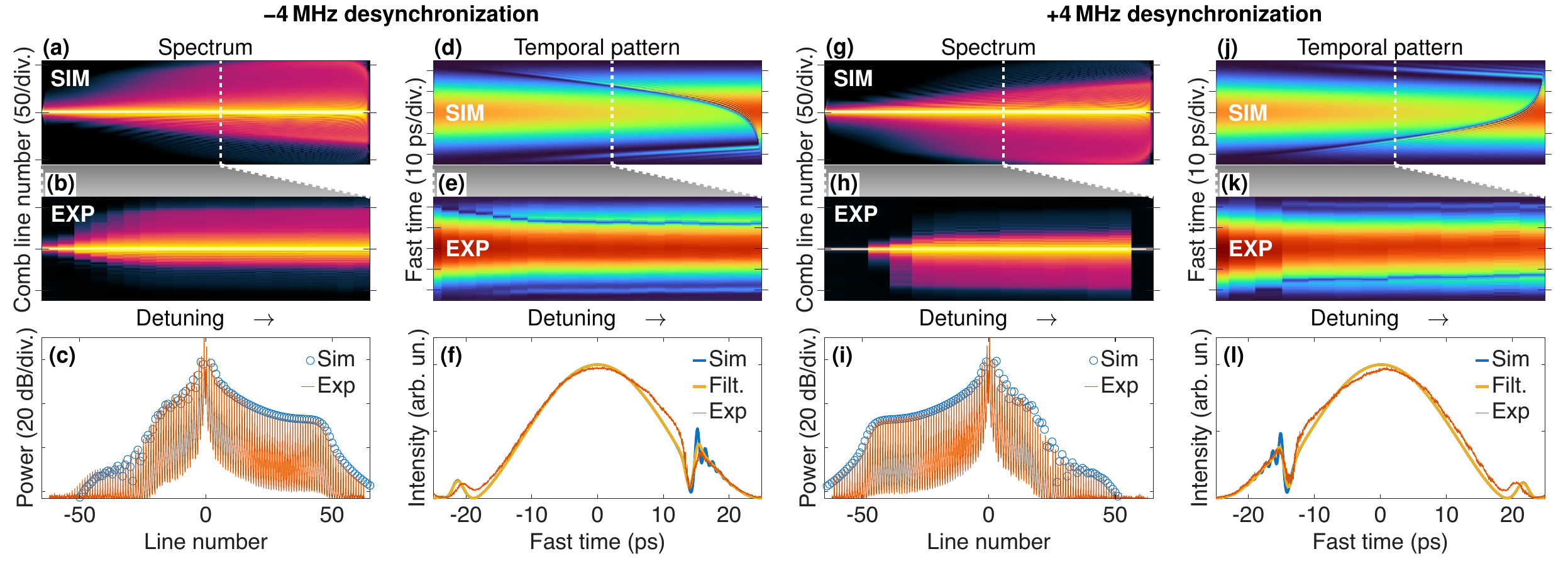}%
	\caption{
		Impact of pump desynchronization on frequency comb dynamics. Panels (a–f) correspond to a negative desynchronization of $\Delta\omega_{\rm sync}/2\pi = \qty{-4}{\MHz}$, while panels (g–l) correspond to a positive desynchronization of $\Delta\omega_{\rm sync}/2\pi = \qty{+4}{\MHz}$.
		\emph{Spectral evolution}: (a, g) Simulated and (b, h) measured evolution of the comb spectral envelope as detuning is scanned over time. (c, i) Direct comparison between experimental and simulated spectra at the maximum achieved detuning.
		\emph{Temporal evolution}: (d, j) Simulated and (e, k) measured evolution of the extra-cavity temporal intensity profile. (f, l) Direct comparison of the experimental and simulated temporal waveforms at maximum detuning.
		The white dashed lines in the simulation maps (a, g, d, j) indicate the points where the experimental scans terminated.
		}
	\label{fig:3}
\end{figure*}

We model the nonlinear dynamic of our system using the Lugiato-Lefever equation (LLE)~\cite{Lugiato1987}
\begin{equation}
	\begin{aligned}
		\frac{\partial A}{\partial t}
		&=
		\left[
		-\left (\frac{\kappa}{2} + i\delta\omega\right )
		+ ig_0 |A|^2
		\right] A\\
		& - i \, \Delta\omega_{\rm sync} \frac{\partial A}{\partial\phi} 
		+ i \sum_{n \geq 2} \frac{D_n}{n!}\frac{\partial^n A}{\partial\phi^n} 
		+ \sqrt{\kappa_{ex}} \, s_{\rm in}(\phi)
	\end{aligned}
	\label{eq:LLE}
\end{equation}
where $A$ is the field envelope, $t$ is the slow time, $\phi \in [-\pi, \pi]$ is the normalized roundtrip time, $\kappa$ is the linewidth accounting for the total energy loss rate, $\delta\omega$ is the detuning, $\kappa_{ex}$ is the external coupling rate, and $g_0$ = $\hbar\omega_0 c n_2/n_g^2V_{\rm eff}$ is the cavity's Kerr nonlinear factor, in which $\hbar$ is the reduced Planck constant, $n_2$ is the nonlinear refractive index, $n_g$ is the effective group index, $V_{\rm eff}$ is the effective mode volume.
In this study, dispersion terms beyond $D_3$ are ignored.
$s_{in}$ is the pump amplitude term. As we use two pump tones separated by one FSR, the pump term in the temporal domain takes the form of a sinusoidal modulation with a single modulation period covering the whole cavity roundtrip $ s_{in}(\phi) = s \, \cos\left( \phi \right) $.
To account for imperfect synchronization between the pump frequency spacing ($2\, f_{mod}$) and the cavity FSR, an additional temporal drift term is added, where $\Delta\omega_{\rm sync} = 2\pi \left( 2\,f_{\rm mod} - \text{FSR}  \right)$ is the desynchronization factor.

To elucidate the origin of the frequency combs observed in our platform, we compare the experimental results with numerical simulations of \cref{eq:LLE} using a split-step Fourier method, where the dual-pump detuning $\delta\omega$ is linearly scanned over time. The simulated spectral envelope and temporal intensity evolutions are presented in \cref{fig:2}a,c. Note that in the temporal map, the \emph{extra-cavity} field is evaluated to replicate the experimental conditions where the OSO records the chip transmission at the through-port of the resonator. At this port, the out-coupled waveform interferes with the residual background pump field that is not coupled into the cavity, governed by:
$s_{\rm out}(\phi) = s_{\rm in}(\phi) - \sqrt{\kappa_{\rm ex}} A(\phi)$. Our direct measurement at the through port, typical of single-coupler microcomb implementations, faithfully reveals the true shape of the out-coupled comb waveform.

The extra-cavity profile is depicted in \cref{fig:2}e at three distinct detuning values. It features a complex envelope characterized by a kink in the sinusoidal pump modulation, flanked by pronounced oscillatory tails that develop with increasing detuning. To uncover the physical origin of these features, it is instructive to examine the corresponding simulated \emph{intra-cavity} intensity profiles (\cref{fig:2}f).
The latter exhibits a more characteristic rectangular profile consisting of high- and low-intensity homogeneous steady states connected via SW~\cite{Coen1999, Parra-Rivas2016}.
As the detuning increases, the system shifts deeper into the bistable regime, elevating the intra-cavity intensity along the upper branch while driving the switching fronts closer together. 
This narrows the flattop pulse width until the fronts eventually collide and mutually annihilate (\cref{fig:2}c). The associated oscillatory tails correspond to the emission of dispersive waves at the boundaries of the fronts~\cite{Coen1999}.
Interestingly, these dispersive-wave features become significantly more pronounced in the extra-cavity field (\cref{fig:2}e) as a direct result of interference with the background pump field, while the fronts appear as kinks.

The dashed white line in the simulated map (\cref{fig:2}a,c) indicates the maximum detuning value achieved experimentally before the laser scan terminated. In practice, the full detuning range predicted by the simulations could not be completely accessed in the experiment.
This limitation is attributed to the absence of active detuning stabilization feedback, leaving the open-loop system vulnerable to ambient environmental and thermal fluctuations that disrupt operation near the extreme edge of the resonance where the intra-cavity power varies rapidly. Despite these practical constraints, a good agreement is observed in both the temporal and spectral maps within this accessible region.

\Cref{fig:2}g compares the simulated and measured temporal profiles of the extra-cavity field at the maximum experimental detuning, demonstrating excellent agreement that validates the numerical model and confirms the profile of the out-coupled waveform. Because the limited bandwidth of the OSO cannot fully resolve the rapid, high-frequency dispersive-wave oscillations at the pulse edges, the simulated data were post-processed using a digital infinite-impulse-response (IIR) Chebyshev type-II filter with an \qty{800}{\GHz} cutoff. This correction yields a highly accurate match with the experimental observations.
As the detuning increases, the temporal narrowing of the pulse naturally drives spectral broadening in the frequency domain (\cref{fig:2}a). At large detunings (\cref{fig:2}h), the generated spectra evolve into characteristic flat-top combs~\cite{Lobanov2015, Liu2022StimulatedGeneration}. The emergence of pronounced spectral shoulders in this regime is directly linked to the development of the oscillatory tails in the time domain~\cite{Macnaughtan2023TemporalCharacteristics, Bunel2024BroadbandKerr}.

In \cref{fig:2}, when the modulation frequency is perfectly synchronized with the FSR of the spiral resonator, both the spectral and temporal profiles remain highly symmetric. However, introducing a mismatch between the modulation frequency and the resonator FSR breaks this symmetry, causing the SW profiles to become asymmetric in both domains~\cite{Liu2022StimulatedGeneration, Bunel2024BroadbandKerr, Macnaughtan2023TemporalCharacteristics}.
Specifically, when the modulation frequency is detuned by $\Delta\omega_{\rm sync}/2\pi = \qty{-4}{\MHz}$, the spectrum shifts toward higher frequencies, exhibiting pronounced asymmetric growth on the high-frequency (blue) side (\cref{fig:3}a-c). This behavior stems from the normal dispersion regime, where higher-frequency components experience a larger group delay and a consequently longer cavity roundtrip time. These slower, high-frequency modes naturally achieve a better phase-matching condition with the reduced modulation frequency, driving the rightward spectral shift.
Conversely, for a positive desynchronization ($\Delta\omega_{\rm sync}/2\pi = +\qty{4}{\MHz}$), preferential spectral growth occurs on the lower-frequency (red) side because those faster modes better match the increased modulation frequency (\cref{fig:3}g-i). This desynchronization-induced asymmetry is equally manifest in the extra-cavity temporal waveforms (\cref{fig:3}d-f,j-l), where the dispersive-wave oscillatory tails become heavily pronounced and shift toward the modulation peak on one side, while being strongly damped and suppressed on the other.

\begin{figure}[htbp]
	\centering%
	\includegraphics[width=\linewidth]{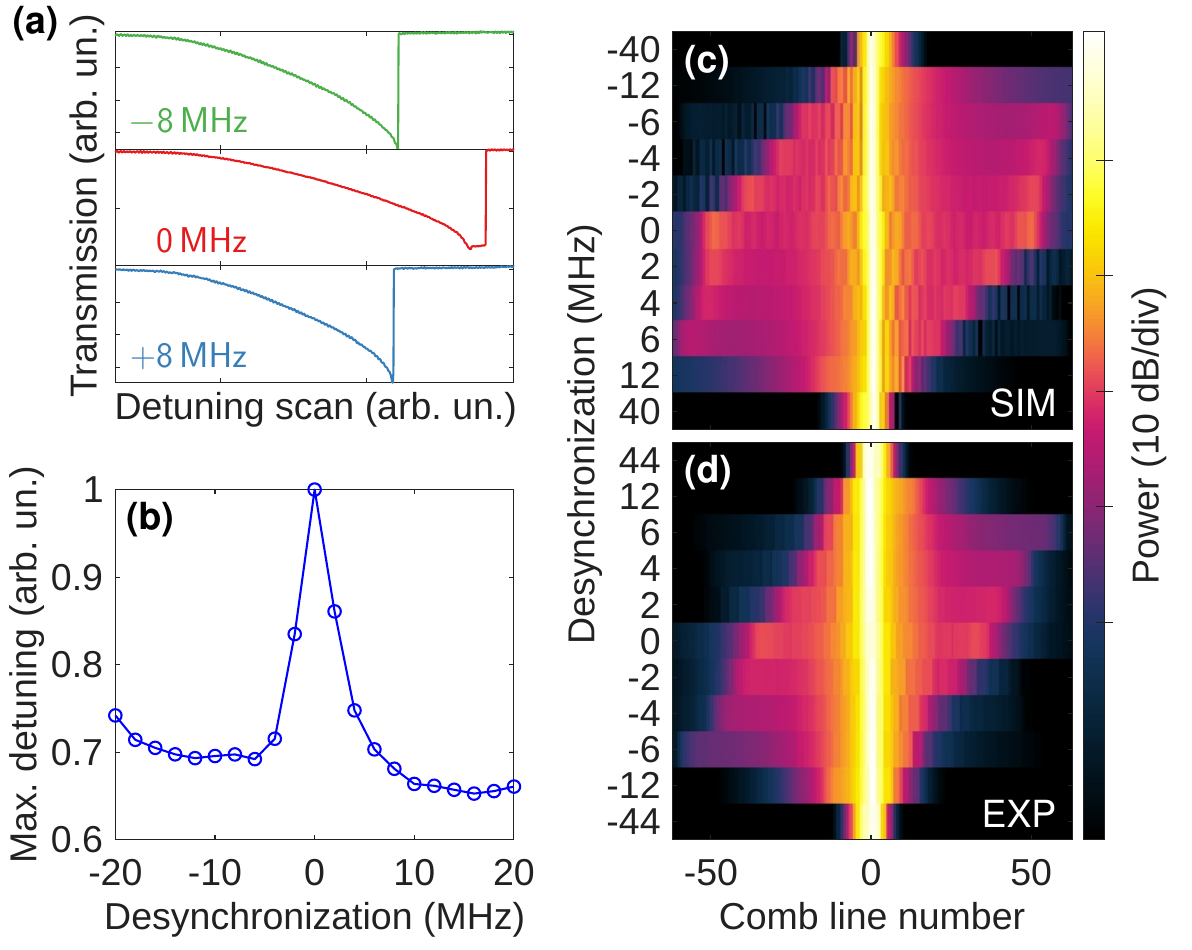}%
	\caption{Detuning and desynchronization.
		(a) Recorded transmission curves at $\Delta\omega_{\rm sync}/2\pi = \qtylist[list-units = single]{-8;0;8}{\MHz}$ .
		(b) Evolution of the maximum detuning relative values when varying desynchronization.
		(c, d) Comparison between the simulated (c) and measured (d) comb spectrum envelope at the maximum detuning, as a function of desynchronization.
	}
	\label{fig:4}
\end{figure}

Next, we systematically explore the interplay between desynchronization, the maximum allowable nonlinear detuning, and the resulting comb asymmetry and bandwidth.
As demonstrated by the measured transmission curves (\cref{fig:4}a), the detuning span of the nonlinear resonance is highly sensitive to the synchronization state.
At zero desynchronization, both SW converge symmetrically toward the peak of the modulation pattern as detuning increases (\cref{fig:2}c), enabling the maximum attainable detuning.
Conversely, introducing a synchronization mismatch (e.g., $\Delta\omega_{\rm sync}/2\pi = \qty{\pm8}{\MHz}$) drastically reduces this detuning span.
Here, the mismatch induces an asymmetric drift, causing one SW to accelerate along the modulation profile with detuning and escape the modulation envelope at a significantly lower detuning value (\cref{fig:3}d,j).
This evolution is quantified in \cref{fig:4}b, which plots the normalized maximum detuning as a function of desynchronization.
While a sharp peak is centered at perfect synchronization, the maximum reachable detuning drops steeply out to $\qty{\pm4}{\MHz}$.
Beyond this threshold, the apparent detuning range widens slightly; however, this is an artifact of the dual-pump scheme. At large desynchronizations, the two pump tones decouple and cross their individual cold cavity resonances at distinct times during the scan.
This offsets their respective transmission dips and broadens the aggregate frequency span, but fails to sustain the simultaneous dual-pump engagement required for a full comb, yielding only narrow, low-efficiency states.
Notably, the subtle asymmetry of this maximum detuning curve between positive and negative desynchronization branches can be attributed to the presence of third-order dispersion.

To evaluate the spectral impact of this behavior, we map the simulated (\cref{fig:4}c) and experimental (\cref{fig:4}d) evolution of the comb envelope at the maximum reachable detuning across a sweep of desynchronization values.
Both maps exhibit a distinctly asymmetric, tilted trapezoidal profile.
This spectral asymmetry peaks near a desynchronization of $\qty{\pm6}{\MHz}$, where the comb achieves its maximum single-sided bandwidth extension.
Beyond this critical threshold, further desynchronization causes a rapid contraction of the comb bandwidth.
In this highly detuned regime, the nonlinearly coupled energy can no longer compensate for the temporal walk-off between the modulation and the cavity roundtrip, precipitating periodic instabilities as the two pump tones lose their simultaneous resonance alignment.

In conclusion, we have demonstrated frequency-comb generation in the normal-dispersion regime using a dual-pumped silicon nitride spiral microresonator. This work provides a robust validation of the compact 'snail-shaped' Archimedean spiral geometry, proving that adiabatic curvature engineering can maintain high quality factors while effectively minimizing avoided mode crossings.

Operating in a strong normal-dispersion and low-repetition-rate regime provided distinct practical advantages in this work. It allows for the creation of combs that have both a high power-per-line budget and a moderate spectral bandwidth that matches that of our OSO.
Crucially, this unique combination of properties allowed for the direct optical sampling of the extra-cavity temporal waveform without the need for complex temporal waveform retrieval method.
The resulting experimental profiles show remarkable agreement with our LLE simulations, confirming the highly predictable SW dynamics of this platform.
Although our choice of strong normal dispersion limited the comb bandwidth, broader comb states can be accomplished by increasing the SiN film thickness to minimize the amount of normal dispersion~\cite{Anderson2020a}.

Looking forward, this successful validation opens the door to more advanced resonator designs. By optimizing the dispersion landscape, this platform is well-positioned to transition toward continuous-wave driven operation. Furthermore, integrating this dispersion-engineered spiral design into a self-injection locking configuration could unlock highly stable, circulator-free, and turn-key integrated microcomb sources, offering a scalable, CMOS-compatible pathway for next-generation photonic integrated circuits.

\section*{Acknowledgments}

This project has received funding from the European Union’s Horizon Europe research and innovation program
under grant agreement No 101137000.

\section*{Competing Interests}

The authors declare no competing interests.
\printbibliography
\end{refsection}

\clearpage

\end{document}